 \documentclass[final,5p,times,twocolumn]{elsarticle}

\usepackage{amssymb}
\usepackage{stfloats}
\usepackage{float}
\usepackage{algorithm}
\usepackage{algorithmic}
\usepackage{amsmath}
\usepackage{amssymb}
\usepackage{color}
\usepackage{multirow}
\usepackage{threeparttable}

\journal{arxiv}

\begin{document}

\begin{frontmatter}

\title{Statistical Feature-based Personal Information Detection in Mobile Network Traffic}
\author[1]{Shuang Zhao}
\ead{zhaoshuang16@nudt.edu.cn}

\author[1]{Shuhui Chen\corref{cor1}}
\ead{shchen@nudt.edu.cn}

\author[1]{Ziling Wei}
\ead{weiziling@nudt.edu.cn}

\cortext[cor1]{Corresponding author}
\affiliation[1]{organization={School of Computer, National University of Defense Technology},
             city={Chang Sha},
             postcode={410073},
             country={China}}

\begin{abstract}
With the popularity of smartphones, mobile applications (apps) have penetrated the daily life of people. Although apps provide rich functionalities, they also access a large amount of personal information simultaneously. As a result, privacy concerns are raised. To understand what personal information the apps collect, many solutions are presented to detect privacy leaks in apps. Recently, the traffic monitoring-based privacy leak detection method has shown promising performance and strong scalability. However, it still has some shortcomings. Firstly, it suffers from detecting the leakage of personal information with obfuscation. Secondly, it cannot discover the privacy leaks of undefined type. Aiming at solving the above problems, a new personal information detection method based on traffic monitoring is proposed in this paper. In this paper, statistical features of personal information are designed to depict the occurrence patterns of personal information in the traffic, including local patterns and global patterns. Then a detector is trained based on machine learning algorithms to discover potential personal information with similar patterns. Since the statistical features are independent of the value and type of personal information, the trained detector is capable of identifying various types of privacy leaks and obfuscated privacy leaks. As far as we know, this is the first work that detects personal information based on statistical features. Finally, the experimental results show that the proposed method could achieve better performance than the state-of-the-art.
\end{abstract}




\end{frontmatter}

%% \newpageafter{abstract}
\section{Introduction}
With the popularity of mobile devices, millions of apps have been developed to facilitate daily living activities. According to FinancesOnline\cite{finonline}, there are 5.22 billion unique mobile phone users worldwide, and a total of 218 billion mobile apps are downloaded in 2020. Apparently, apps have been indispensable in the daily life of people. 

When providing services, apps would request access to personal information (PI), such as device identifiers and location\cite{chadza2017,He2019}. Some of the requested PI is necessary to support the functionalities of apps, while some PI is deliberately collected by apps with unilateral intent. Since the majority of users lack professional knowledge, they will unconsciously assent to all requests of the app, even if some requests are unreasonable. In this case, users would be victims of unnecessary privacy leaks. For example, PI of up to 87 million Facebook users had been collected by a third-party Cambridge Analytica without user awareness\cite{facebook}. Cambridge Analytica developed a quiz app. Once a user logs into the quiz app with his Facebook account, his data could be collected by the quiz app, such as profile information, user history, the list of friends, etc. Then the data can be used to automatically predict a range of highly sensitive personal attributes, such as sexual orientation, personal traits, etc. Similarly, Rela, a Chinese social app, was found to have leaked 5.3 million user profiles\cite{Rela}. With the disclosure of such privacy leakage events, more and more users are aware of privacy concerns and are likely to appreciate transparency when regarding how the app collects personal data. Furthermore, after investigating how privacy leaks change over time for 512 apps, Ren et al.\cite{Ren2018} point out that the data leakage of apps has gotten worse. Therefore, there is an urgent need to detect privacy leaks in apps. 

Existing privacy leak detection methods can be divided into three categories, including static analysis-based detection\cite{Egele2011,AndroidLeaks}, dynamic analysis-based detection\cite{MobileappScrutinator,AGRIGENTO}, and network traffic monitoring-based detection\cite{Antmonitor2015, Recon2016, Antshield2018}. With data flow analysis technology, the static analysis-based method finds the privacy leakage path from the source code of an app. Such a method is scalable, while its accuracy is heavily affected by false positives and false negatives. Besides, the effectiveness of this method will further decline with the popularization of source code protection technology. The dynamic analysis-based method usually utilizes customized OS combined with taint analysis to discover the privacy leakage behavior. It could achieve better accuracy than the static analysis-based methods. However, it is difficult to deploy such a method on a large scale due to the high overhead. Compared with the above two methods, the network traffic monitoring-based method provides a user-friendly and scalable detection scheme. This method first monitors the traffic of apps passively. Then it analyzes the content of the traffic and discovers the PI that is transmitted through the traffic. For instance, Recon\cite{Recon2016} regards the traffic as a short text and applies the bag-of-words model to extract features. Then a Random Forest classifier is trained to detect the personally identifiable information (PII) in the traffic. The network traffic monitoring-based method can be easily deployed on personal mobile devices. Besides, its detection performance is comparable to those of the other two methods. Hence, this method has aroused considerable interest among related parties\cite{zhang2021}. 

However, most of the existing traffic monitoring-based detection methods identify PI based on the semantics and formats of PI, which leads to some shortcomings. On the one hand, those methods have poor performance in detecting the PI with obfuscation, such as the PI that is hashed or encrypted. On the other hand, they cannot identify the PI of undefined types (e.g.PI with unknown formats). In this paper, a novel traffic monitoring-based detection method is proposed. It could solve the above two problems effectively. Instead of distinguishing PI and non-PI by their semantic and formats, our method focuses on the differences in occurrence patterns between them. Specifically, our method is inspired by the observation that PI and non-PI transmitted in the traffic could show different occurrence patterns. Take IMEI (a device identifier) and the timestamp (a non-PI) as examples. An app would obtain IMEIs from multiple users multiple times, while the IMEIs from the same user are unique. Meanwhile, the IMEI of a user would be collected by multiple apps, hence the IMEI could occur in the traffic of multiple apps. As for the timestamp, timestamps collected from one user would change frequently and a timestamp rarely appears in multiple apps at the same time. For PI and non-PI, their occurrence patterns are independent of their semantics and formats. Therefore, the undefined PI can be discovered as long as it has similar occurrence patterns. In addition, the obfuscated PI can also be detected by our method when the obfuscation does not affect the patterns of how PI occurs in the traffic.

The occurrence patterns are captured from local and global perspectives in our method. Specifically, local and global statistical features are designed to describe how PI occurs in the traffic of one app and all apps. Then a group of string-form rules and appropriate manual work are employed to build a labeled dataset. Finally, a detector is trained to learn the occurrence patterns of PI and non-PI based on machine learning algorithms. With the detector, more potential PI in traffic can be discovered automatically. The contributions of this paper are summarized as follows:
\begin{enumerate}[(1)]
\item A set of local and global statistical features is designed to describe the occurrence patterns of PI in the traffic. To our best knowledge, this is the first work to detect PI in the traffic from the perspective of statistical features.
\item An PI detector based on machine learning is presented in this paper. The detector could identify various potential PI without knowing the format and semantics of PI. Besides, the detector is effective for detecting a portion of obfuscated PI.
\item Comprehensive experiments are conducted on a real-world large-scale dataset to validate the effectiveness of the proposed detector. The experimental results show that the proposed detector could achieve better performance than the state-of-the-art.
\end{enumerate}

The rest of the paper is organized as follows. Section 2 provides the necessary background and introduces the related work. Then designed statistical features are given in Section 3. Section 4 presents the detection method. Section 5 shows the experimental results. Finally, Section 6 concludes the paper.

\section{Background}
In this Section, the detailed explanations of two terms (i.e. PI and privacy leak) used in this paper are provided first. Then, the related work of three types of privacy leak detection methods is introduced. 
\subsection{Personal Information}
PI refers to all kinds of information that can identify a specific individual or reflect the activities of an individual\cite{GB2020}. Under different standards and application scenarios, the specific information belonging to PI is different. PII refers to the PI that can be used to identify or track an individual, which is the most sensitive information of a person. So far, there are several types of well-defined PII, including device identifiers, user identifiers, contact information, credentials, location. Current researches basically focus on identifying those PII. However, in addition to those PII, there are other undefined PII and PI that can reveal the sensitive information of users. Therefore, it is far from enough to identify only the well-defined PII.

Our work aims at identifying as much PI as possible. A critical problem is how to define those PI, such as what they look like and what data they transmit. Fortunately, our work could neatly bypass this problem. During the detection, the proposed method in this paper analyzes the data in the traffic based on the statistical features. Other knowledge, such as APIs for accessing the PI and the format of the PI, is not required. In this way, the proposed method could discover the PI without a precise definition of the PI. 

\subsection{Privacy Leak}
Term privacy leak has been widely used to represent the behavior of apps collecting and transmitting any PI\cite{AndroidLeaks,Recon2016}. Recently, a few studies further assess the rationality of such behaviors. In other words, those studies try to figure out whether the behavior is necessary for the app to provide its service\cite{LeakDoctor,BATLeaks}. In those studies, privacy leak indicates the functionality-irrelevant private data transmission. 

Since the rationality of transmitting PI is beyond the scope of this paper, privacy leak represents any PI transmission behavior in this paper. That is any traffic flow that transmits PI is regarded as a privacy leak. 

\subsection{Related Work}
\subsubsection{Static Analysis-based Detection}
The static analysis-based detection method discovers the possible privacy leak in an app by analyzing its source code. The first step of this method is to find the source and sink in the source code. For example, the APIs provided by the device OS for accessing user data can be regarded as sources. The interfaces that transmit information are defined as sinks, such as network interface and file writing functions. Then the control flow graph of the source code is generated. Finally, data flow analysis technology is utilized to find paths between the sources and sinks. Each path indicates a potential privacy leak. 

Since the static analysis is carried out without running apps, it has good scalability. Thus it can be applied to implement the preliminary test of massive apps in the application market. However, this method is prone to generate false negatives. On the one hand, it is hard to analyze the native code and dynamically loaded code. On the other hand, the privacy leaks generated from non-standard sources (e.g. PI input by users) would be omitted. Besides, since this method finds the leakage paths statically, false positives would be introduced if the paths are never visited. 

Based on static analysis, FlowDroid\cite{FlowDroid2014} finds the leakage paths between the predefined sources and sinks. Apart from that, it proposes a scheme to find the privacy leak caused by the UI widgets within the apps. Nattanon et al.\cite{Nattanon2020} extend FlowDroid by adding additional PII-related sources into its source-sink file. Besides, a PII list assembled from previous studies is provided in their work. AndroidLeaks\cite{AndroidLeaks} provides a static analysis framework for discovering PII leaks in Android apps. PiOS\cite{Egele2011} analyzes the binaries compiled from Objective-C code and detects privacy leaks in iOS apps. ClueFinder\cite{ClueFinder} also performs its detection based on the source code of apps. Instead of using data flow analysis, it detects the PI leakage paths by finding the program elements with PI semantics.

\subsubsection{Dynamic Analysis-based Detection}
The dynamic analysis-based detection performs its data flow analysis with running apps. It could achieve better performance compared with the static analysis since the discovered privacy leaks happen in real data streams. To track how personal data flows in the app, some detection methods apply taint analysis on customized OSs. TaintDroid\cite{TaintDroid2014} modifies the Android OS to support attaching taint tags to sources. Four granularities of taint propagation are implemented in TaintDroid, including variable-level, method-level, message-level, and file-level. Achara et al.\cite{MobileappScrutinator} present MobileappScrutinator, a dynamic analysis platform for Android and iOS. MobileappScrutinator rewrites the source code of Android OS to track personal data. Then similar functions are implemented in iOS with jailbreaking. PrivacyCapsules\cite{privacycapsules} implements a customized OS. It requires the apps running above it to comply with its PI access rules. At the same time, the PI access rules prevent the PI from leaving the personal device. iABC\cite{iABC} evaluates the risks of privacy leaks of iOS apps by combining dynamic analysis with static analysis. With a customized OS, He et al.\cite{He2018} hook the privacy-related APIs to identify privacy leaks of the third-party libraries inside the apps. However, these methods usually run the apps with automatic tools, which leads to incomplete coverage of app execution paths\cite{Ren2018}. Therefore, false negatives would be introduced. Meanwhile, those methods cannot eliminate false positives because of coarse-grain taint and tainted information explosion\cite{TaintDroid2014}. 

Another method is proposed based on differential analysis. AGRIGENTO\cite{AGRIGENTO} first establishes the network behavior baseline of an app by running the app multiple times. Then the input value of the PI to the app is changed. Finally, the privacy leaks are found by observing deviations in the resulting network traffic. In this way, AGRIGENTO also could detect obfuscated PI in the traffic.  Although this method could achieve high precision, its practicability is relatively poor as a large amount of manual operation is involved.

\subsubsection{Traffic Monitoring-based Detection}
The traffic monitoring-based detection passively monitors all the traffic generated by the apps, then it detects the transmitted PI in the traffic. The traffic is generally monitored by self-developed tools\cite{Antmonitor2015,Recon2016,Antshield2018,PrivacyGuard,Haystack}. These tools utilize the VPNService provided in Android OS to collect the traffic without requiring root access. Besides, since these tools are deployed on mobile devices, the source app of the traffic could be obtained simultaneously. Compared with the other two types of detection methods, the traffic monitoring-based detection method is lighter and easier to deploy. In addition, it could continuously monitor the traffic generated by apps hence full coverage of detection can be achieved.  

With the self-developed tool Antmonitor\cite{Antmonitor2015}, Anastasia et al. find the PI that is readily available to apps on the phone use simple string matching, such as IMEI, email, and phone number. PrivacyGuard\cite{PrivacyGuard} adopts regex to detect the PI with specific formats in the traffic. Similarly, Liu et al.\cite{Liu2015}. design regex to find five kinds of PI in the traffic. Then the values of the discovered PI are added to the regex rules to mine more privacy leaks. Recon\cite{Recon2016} firstly proposes a detection scheme based on machine learning. It regards the traffic flows as short texts. Then the bag-of-words model is applied to extract the features of the traffic flows. Finally, a decision tree classifier is trained for each domain to identify five types of PII. Anastasia et al.\cite{Antshield2018} present Antshield, a similar on-device privacy leaks detection method. Besides finding the known PII based on string matching, Antshield detects the unknown PII by machine-learning classifiers, which are similar to Recon. Compared with Recon, Antshield trains a classifier for each app instead of for each domain. 

A weakness of the traffic monitoring-based detection methods is that they cannot deal with encrypted traffic unless extra technology is adopted, such as man-in-the-middle (MITM) proxy. The method proposed in this paper also suffers from this problem. Additionally, the detection methods based on simple string/regex matching are not resilient to obfuscation technology. Lastly, the existing methods can only detect the PII with predefined types, while our method takes a step toward the detection of undefined PI.

\section{Personal Information Features}
In this Section, $<app, \ Key>$ pair, which is the basic detection object of our method, is first defined in Section 3.1. Then the designed local features and global features for a $<app, \ Key>$ pair are presented in Section 3.2. Those features are used to depict the occurrence patterns of a $<app, \ Key>$ pair.

\subsection{$<app, \ Key>$ Pair}
\begin{figure}[h]
    \scriptsize
    \centering
    \includegraphics[width=0.3\textwidth]{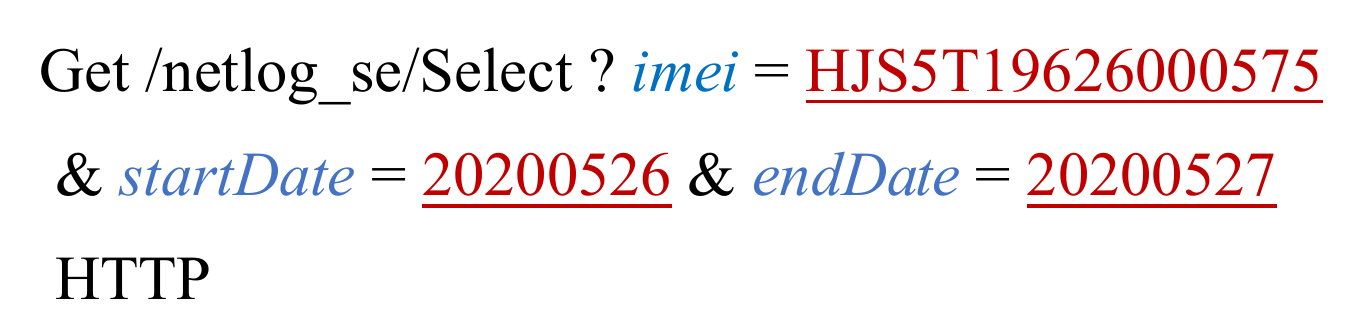}
    \caption{An example of a HTTP request from Wechat. The keys are marked in blue italics. The values are marked in red and underlined. The key and the value in one $<key, \ value>$ pair are connected with '=' in the HTTP request.}
    \label{fig.1}
\end{figure}

For the unencrypted mobile traffic, most of the PI is transmitted through $<key, \ value>$ pairs in the HTTP request\cite{AGRIGENTO,Liu2015}. Fig.\ref{fig.1} gives an example of a HTTP request from WeChat, one of the most popular instant chatting app in China. As Fig.\ref{fig.1} shows, there are three $<key, \ value>$ pairs in the request, i.e, $<imei, \ HJS5T19626000575>$, $<startDate, \ 20200526>$, and $<endDate, \  20200527>$. 

In our previous work\cite{Shen2019}, it has been observed that HTTP requests generated by the same action in an app are likely to have a similar structure. More precisely, those HTTP requests have the same keys while the values may change. Besides, considering that the developers of one app usually adopt certain naming conventions to name keys, we further assume that a key in an app constantly transmits the same type of information. Based on this assumption, the detection task is transformed into identifying whether each $<app, \ Key>$ pair is PI-related. Once it is determined that pair $<app_i, \ key_j>$ transmits PI, then a privacy leak occurs when $key_j$ appears in the HTTP request of $app_i$ and the value of $key_j$ is not empty or a default value.

\subsection{Statistical Features of $<app,Key>$ Pairs}
For clarity, the symbols used to define the proposed features are first given as follows. Suppose that the traffic set contains \textit{n} apps. The app set is denoted as $A=\{app_1,...,app_n\}$. For pair $<app_i, \ key_j>$, the different values of $key_j$ compose $V_{ij}=\{v_{ij}^1,v_{ij}^2,...,v_{ij}^m\}$, $\#(v_{ij}^{t})$ calculates how many times $v_{ij}^{t}$ appears. $H_{ij}$ denotes the set of domains that visited by the $app_i$’s HTTP requests containing $key_j$. $V_{ij}$ and $H_{ij}$ are regards as the attributes of $key_j$. In following sections, 11 local features and 6 global features are designed to characterize the occurrence pattern of $<app, \ Key>$ pairs. For pair $<app_i, \ key_j>$, local and global indicates whether the features are extracted from the traffic of $app_i$ or app set \textit{A}.

\subsubsection{local features}
For pair $<app_i, \ key_j>$, its local features include the statistical features of $key_j$ that extracted from the traffic of $app_i$. 

(1) Key-value statistics

For some PI, such as device identifiers, their values for a user would be unique or remain unchanged for a period of time. In contrast, the value of the non-PI may change frequently, such as the visited resource, timestamp, etc. This difference is described by two metrics in our method, i.e., the number of different key values per user and the value entropy per user. For pair $<app_i, \ key_j>$, the above two metrics are first computed for each user. Since different $<app, \ Key>$ pairs may have a different number of users, 8 statistical features are extracted as key-value statistics, including max/min/ava/var of the results of the two metrics. In a word, the key-value statistics indicate how the values of one user on $key_j$ change in the traffic of $app_i$.  

(2) Local-Value Reuse Degree (L-VRD)

L-VRD is the ratio of the values in $V_{ij}$ used by at least two users. This feature implies the coincidence of values taken by users.

(3) Key Frequency

Key frequency refers to the proportion of HTTP requests that contain $key_j$ to all HTTP requests from the $app_i$, i.e., the frequency of $app_i$ uses $key_j$ to obtain the information. This feature is designed to describe the difference in the frequency at which PI and non-PI are collected by apps. 

(4) Number of Users

Lastly, to reduce the impact of the number of users on the above features, the number of users that contribute to the traffic of $app_i$ is added as one of the local features.

\subsubsection{global features}
When the occurrence pattern is inferred only by local features, it is prone to introduce false predictions. Take pair $<Wechat, \ appver>$ as an example. The \textit{appver} is used to carry the app version of the Wechat installed on the device, which is not PI. However, $<Wechat,\ appver>$ could show similar local features as the PI-related pairs when the traffic of Wechat is collected from one user. Therefore, global features are further designed to characterize how $key_j$ and its attributes distribute in the traffic of all apps.

(1) Key Reuse Degree (KRD)
 
KRD refers to the number of other apps that take $key_j$ as one of their keys. Its calculation is defined in Eq.\ref{eq.1} and Eq.\ref{eq.2}. KRD indicates whether $key_j$ is app-specific or is commonly employed among apps. 
\begin{equation}
KRD=\sum_{k=1,k\neq i}^{n}f(key_j,app_k) \label{eq.1}
\end{equation}

\begin{equation}
f(k,app)=\left\{\begin{matrix}  \label{eq.2}
1, & if\ k\ is\ one\ of\ the\ keys\ of\ app.\\ 
0, & else
\end{matrix}\right.
\end{equation}

(2) Domain Reuse Degree (DRD)

DRD refers to the number of other apps that visit at least one domain in $H_{ij}$, as shown in Eq.\ref{eq.3} and Eq.\ref{eq.4}. If $H_{ij}$ is visited by multiple apps, it is possible that $key_j$ is generated by a third-party library rather than the function of $app_i$ itself. DRD is designed to distinguish the patterns of how the third-party library collects information from that of the app itself.
\begin{equation}
DRD=\sum_{k=1,k\neq i}^{n}g(H_{ij},app_k) \label{eq.3}
\end{equation}

\begin{equation}
g(H,app)=\left\{\begin{matrix} \label{eq.4}
1, & \exists h \in H_{ij},\ app\ visits\ h.\\ 
0, & else
\end{matrix}\right.
\end{equation}

(3) Weighted Value Distribution Features

\begin{figure}[h]
    \scriptsize
    \centering
    \includegraphics[width=0.3\textwidth]{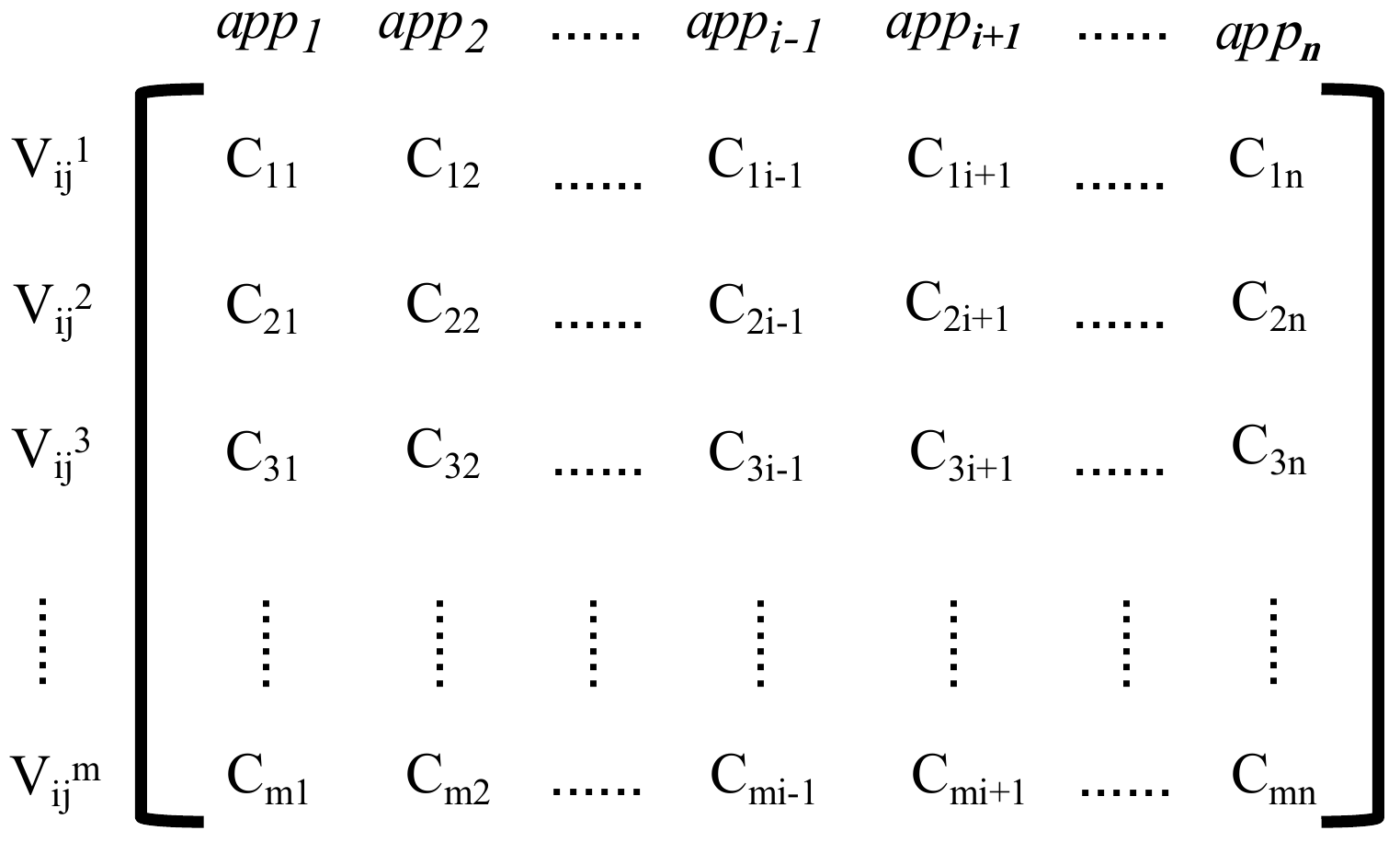}
    \caption{The Value Distribution Matrix of $<app_i, \ key_j>$.}
    \label{fig.2}
\end{figure}

The weighted value distribution features reflect the distribution of $V_{ij}$ in the traffic of other apps. To provide formal descriptions to these features, Value Distribution Matrix (VDM) is defined in our work. Fig.\ref{fig.2} illustrates the VDM of pair $<app_i, \ key_j>$, a matrix with a size of  m*(n-1). Element $C_{tk}$ is one kind of statistical data that relates to the occurrence of $v_{ij}^{t}$ in the traffic of $app_k$. Besides, $W_{ij}=\{w_{ij}^1,w_{ij}^2,...,w_{ij}^m\}$ is the weight vector of $V_{ij}$, $w_{ij}^{t}$ is the weight of $v_{ij}^{t}$, as computed in Eq.\ref{eq.5}.

\begin{equation}
w_{ij}^t=\frac{\#(v_{ij}^t)}{\sum_{t=1}^{m}\#(v_{ij}^t)} \label{eq.5}
\end{equation}

\begin{figure*}[b]
    \scriptsize
    \centering
    \includegraphics[width=0.75\textwidth]{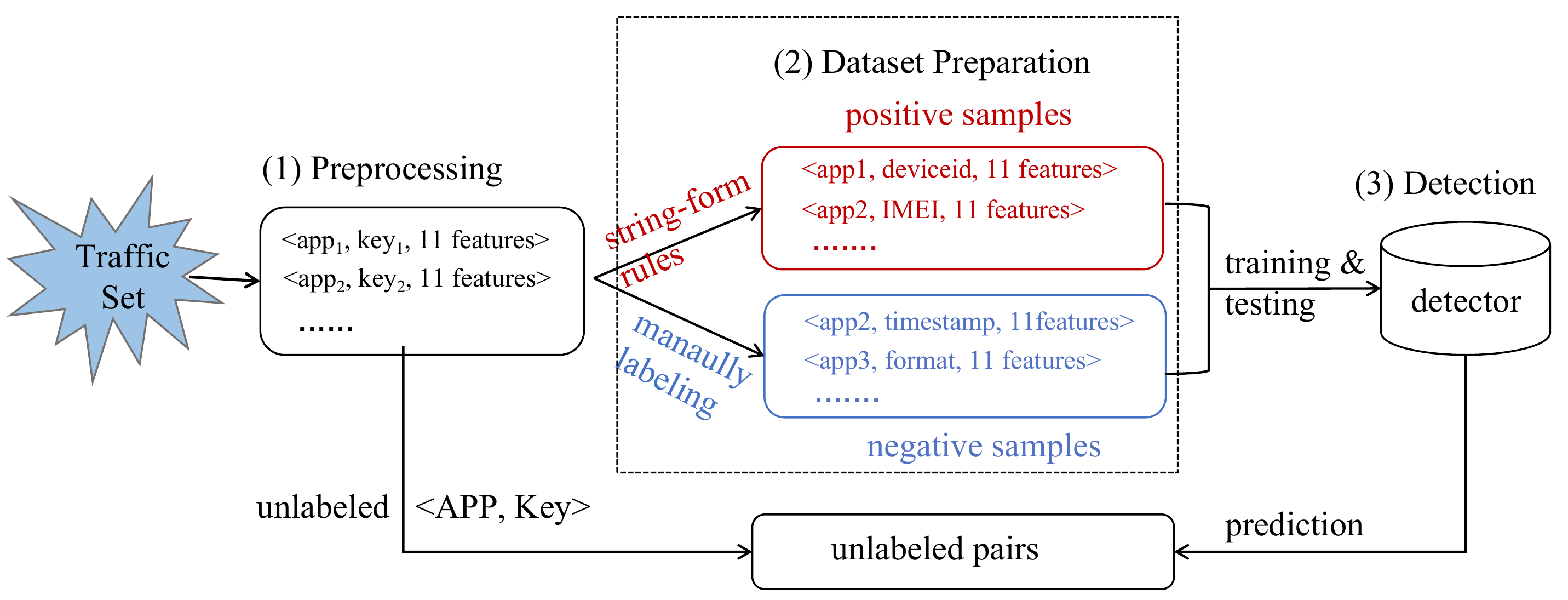}
    \caption{The implementation of the detection method.}
    \label{fig.3}
\end{figure*}

Based on four different $C_{tk}$, four global features are designed as follows: 
\begin{itemize}
  \item Weighted Global-Value Reuse Degree (Weighted G-VRD)

   The weighted G-VRD is the weighted sum of the number of occurrences of the values in $V_{ij}$ in other apps' traffic. Its calculation is given in Eq.\ref{eq.6}, where $C_{tk}$ is the number of times $v_{ij}^t$ appears as a value in the traffic of $app_k$.
   \begin{equation}
   weighted\ GVRD=\sum_{t=1}^{m}(w_{ij}^t*\sum_{k=1,k \neq i}^{n}C_{tk}) \label{eq.6}
  \end{equation}

  \item Weighted app Reuse Degree (Weighted ARD)
 
  The weighted ARD is similar to the weighted G-VRD, except for $C_{tk}$ indicates whether $v_{ij}^t$ appears in the traffic of $app_k$. In other words, $C_{tk}$ is 1 when $v_{ij}^t$ appears in the traffic of $app_k$; otherwise, $C_{tk}$ is 0. 

  \item Weighted User Reuse Degree (Weighted URD)

  The weighted URD is the weighted sum of the number of users using $v_{ij}^t$ in the traffic of other apps.  As Eq.\ref{eq.7} shows, $C_{tk}$ is the user set using value $v_{ij}^t$ in the traffic of $app_k$.
   \begin{equation}
   weighted\ URD=\sum_{t=1}^{m}(w_{ij}^t*\left | \bigcup_{k=1, k \neq i}^{n}C_{tk} \right |) \label{eq.7}
  \end{equation}

  \item Weighted New User Reuse Degree (Weighted NURD)
   Compared with the weighted URD, the weighted NURD concerns new users. Suppose the users involved in pair $<app_i, \ key_j>$ is denoted as U, the weighted NURD of pair $<app_i, \ key_j>$ is calculated as Eq.\ref{eq.8}. The weighted NURD indicates how many new users use the values in $V_{ij}$ in other apps.
  \begin{equation}
   weighted\ NURD=\sum_{t=1}^{m}(w_{ij}^t*\left | (\bigcup_{k=1, k \neq i}^{n}C_{tk}) - U\right |) \label{eq.8}
  \end{equation}
\end{itemize}

\begin{table*}[t]
\scriptsize
\renewcommand\arraystretch{1.3} 
\setlength\tabcolsep{11pt}
\centering
\caption{The rules for positive samples discovering}
\label{tab1}
\centering
\begin{tabular}{cc}
\hline
PI Type           & String-form Rules                                                                                                             \\ \hline
User Identifier   & user, userid, user\_cid, user\_id, user-id                                                                               \\ 
Device Identifier & \begin{tabular}[c]{@{}c@{}}imei, meid, imsi, misi, deviceid, device\_id, serialnumber\end{tabular}                    \\ 
MAC Address       & \begin{tabular}[c]{@{}c@{}}mac, mac\_address, ((([a-f0-9A-F]\{2\}:)\{5\}z)$\mid$(([a-f0-9A-F]\{2\}-)\{5\}))[a-f0-9A-F]\{2\}\end{tabular}                                                       \\ 
Location          & \begin{tabular}[c]{@{}c@{}}location,  gps, latlng, longitude, ltt, lat, latitude, lgt, lng, lon, address\end{tabular} \\ 
Email             & [a-zA-Z0-9\_.+-]+@[a-zA-Z0-9-]+\.[a-zA-Z0-9-.]+                                                                                                                    \\
Phone Number      &  \^{}(86)?(13[0-9]$\vert$14[5$\vert$7]$\vert$15[0-9]$\vert$166$\vert$17[3$\vert$6$\vert$7$\vert$8]$\vert$18[0-9]$\vert$19[1$\vert$8$\vert$9])\verb|\|d\{8\}\$  \\ \hline
\end{tabular}
\end{table*}

\section{Detector Construction}
Based on the proposed statistical features, the PI detection method proposed in this paper is implemented as Fig.\ref{fig.3} demonstrates. Suppose that a traffic set that is marked with source apps has been collected. At the preprocessing stage, $<app, \ Key>$ pairs and their features are extracted from the traffic. To construct a labeled dataset, a group of string-form rules is applied to discover a small number of positive samples, i.e, the $<app, \ Key>$ pairs that transmit PI. Meanwhile, negative samples are labeled with appropriate manual work. Finally, a binary-classification detector is trained based on machine learning. It can be applied to identify whether the unlabeled samples are PI-related.

\subsection{Preprocessing}
In this stage, all $<app, \ Key>$ pairs are extracted from the HTTP requests. Then the $<app, \ Key>$ pair is removed if all of its values are default or empty. The default values are collected by examining a part of sampled traffic, including `none', `unknown', `-', `[IMEI]', `[MAC]', etc. Besides, the $<app, \ Key>$ pairs whose key appears only once in the HTTP requests are also removed. For those pairs, their data volume is too small to represent meaningful occurrence patterns. At last, 11 statistical features are extracted for each valid $<app, \ Key>$ pair. Note that the default values are deleted from the value set of $<app, \ Key>$ pairs when extracting features.

\subsection{Dataset preparation}
To train the detector, a labeled dataset is required. Since we have no prior knowledge of PI, the samples cannot be labeled accurately based on the values of the PI. In this paper, string-form rules and manual labeling are employed to find a handful of positive samples and negative samples, respectively. 

The string-form rules include a group of keywords and regex for six kinds of PI. The rules are listed in Table \ref{tab1}. The keywords come from our previous work\cite{ours2019}, which summarizes the commonly used keywords that transmit PI. In addition, regex rules are applied to find MAC address, email address, and phone number since these PI have specific formats. Based on these rules, some positive samples can be discovered. More specifically, for pair $<app_i, \ key_j>$, if $key_j$ is one of the keywords or at least one value of $key_j$ satisfies the regex, it is labeled as positive. Furthermore, the values of the found positive samples can be exploited to discover more positive samples. Finally, the found positive samples are manually checked and the unqualified samples are deleted. It is worth mentioning that these rules can be flexibly modified and expanded in different scenarios.

As for the negative samples, they are found and labeled manually in our work. On the one hand, the negative samples lack prominent features to be discovered automatically. On the other hand, only a small quantity of positive samples can be found by the string-form rules. Hence, the labor work introduced by manually labeling the same quantity of negative samples is completely acceptable. Furthermore, thanks to a large number of $<app, \ Key>$ pairs are non-PI related, negative samples can be labeled easily in practice. Although the final labeled dataset has a small size, the experimental results in Section 5 show that it is enough to train an effective detector.

\subsection{Detector}
Finally, a detector is trained based on the machine learning algorithm to predict whether a $<app, \ Key>$ pair transmits PI. As previously stated, the proposed detector in this paper can identify unknown types of privacy leaks. In other words, although our labeled dataset contains only six types of PI-related samples, the trained detector could identify the $<app, \ Key>$ pairs that transmit PI beyond those six types. 

From the implementation of the detector, it can be seen that the above detection can be performed after a certain amount of traffic has been collected. However, the advantage of this detection model is that it could predict a large number of $<app, \ Key>$ pairs within one prediction. Then a blacklist can be built from the $<app, \ Key>$ pairs predicted as positive. After deploying the blacklist in the traffic monitoring apps similar to Recon\cite{Recon2016}, only simple string matching is required to find the privacy leaks in real-time traffic. While processing the real-time traffic, new $<app, \ Key>$ pairs and their data can be sent to a centralized server so that the detector and its prediction results can be updated periodically. 

\section{Experiments}
\subsection{Dataset}
In this paper, a self-developed app Netlog\cite{Netlog} is utilized to collect the traffic of mobile devices. The working principle of Netlog is similar to the tool used in Recon\cite{Recon2016}. Then 224 volunteers are recruited to participate in the collection of app traffic.  At last, the collected traffic dataset covers 350 apps with a total of 5,374,633 HTTP requests. This dataset is planned to be made public in our future work.

 \begin{figure}[h]
    \scriptsize
    \centering
    \includegraphics[width=0.4\textwidth]{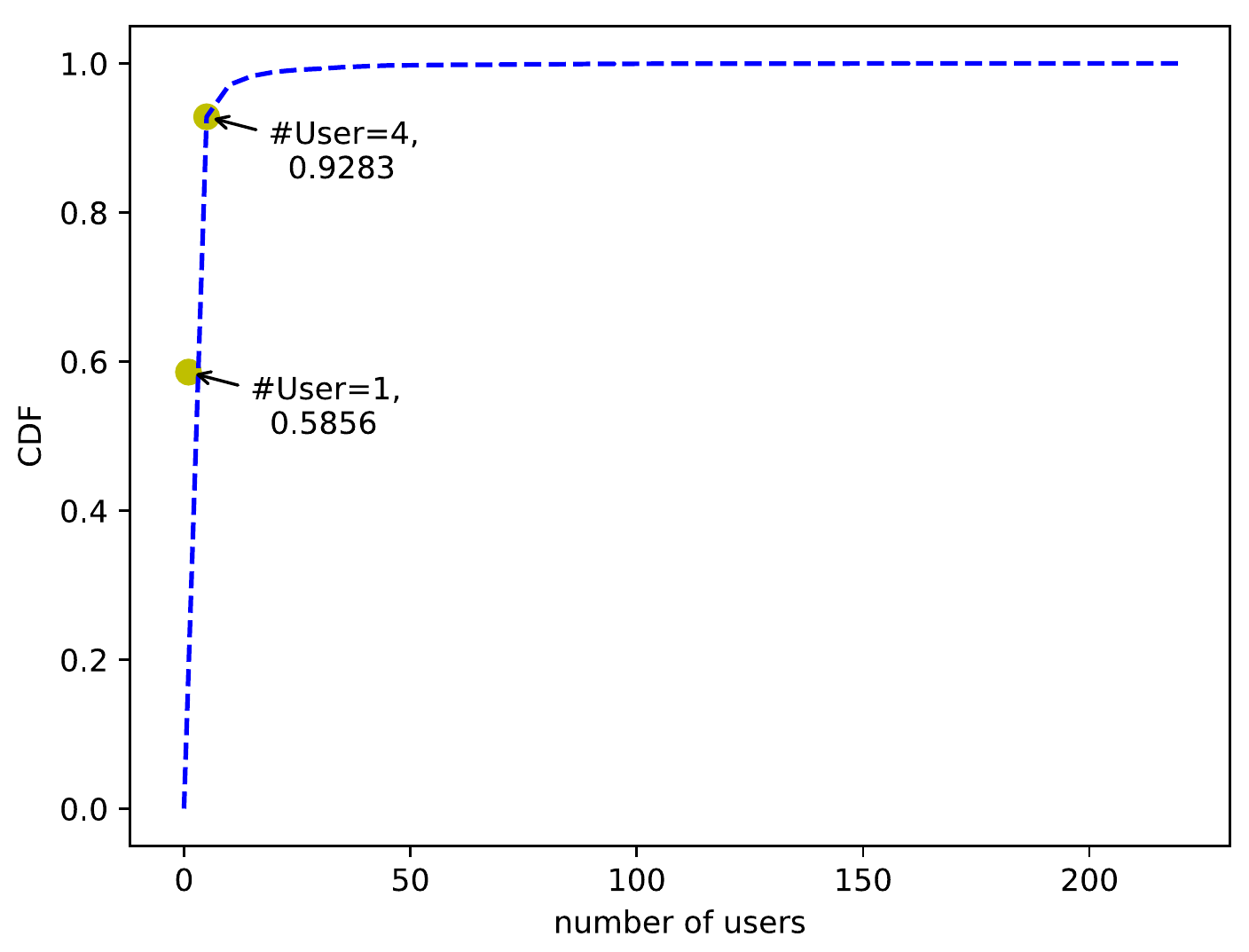}
    \caption{The CDF of the number of users in $<app, \ Key>$ pairs.}
    \label{fig.4}
\end{figure}

After preprocessing, 16,968 valid $<app,\ Key>$ pairs are extracted from 335 apps. The Cumulative Distribution Function (CDF) of the number of users in $<app, Key>$ pairs is depicted in Fig.\ref{fig.4}. It can be seen from Fig.\ref{fig.4} that more than 90\% of the $<app,\ Key>$ pairs have less than five users in our dataset. Meanwhile, 58.56\% of the $<app, Key>$ pairs only appear in the traffic of one user. In other words, more than half of the $<app, Key>$ pairs do not have values from different users, which further means that they cannot obtain reliable local features. Consequently, the difficulty of correctly classifying such samples would increase. Subsequent experimental results further prove this viewpoint.

Based on the labeling process introduced in Section 4.2, 404 positive samples and 404 negative samples are obtained from the 16,968 samples. Similar to the user distribution of the whole dataset, 59.6\% of the 808 labeled samples have only one user, and 94.5\% of the samples have no more than four users. The composition of positive samples is listed in Table \ref{tab2}. 

\begin{table}[h]
\scriptsize
\renewcommand\arraystretch{1.3} 
\setlength\tabcolsep{7pt}
\centering
\caption{The composition of positive samples.}
\label{tab2}
\begin{tabular}{cccc}
\hline
PI       & \# of \textless{}app, Key\textgreater{} & \# of Leaks & \# of apps \\ \hline
Email    & 2                                       & 438         & 2          \\
IMEI     & 124                                     & 556,389     & 67         \\
Phone Number    & 9                                       & 49,453      & 7          \\
other DeviceID & 92                                      & 18,340      & 76         \\
Location & 64                                      & 22,516      & 34         \\
MAC      & 66                                      & 21,901      & 57         \\
UserID   & 47                                      & 134,626     & 41         \\
Total    & 404                                     & 803,663     & 162        \\ \hline
\end{tabular}
\end{table}

Several observations can be made from Table \ref{tab2}. 
First of all, the leakage of IMEI is the worst. About 20\% of the apps transmit IMEI in 10.35\% of the HTTP requests. This should be taken seriously since IMEI can be directly used to identify an individual. Secondly, although 47 $<app,\ Key>$ pairs transmit UserID, which is less than that of the other deviceID and MAC, the leakage times of UserID exceed that of other deviceID and MAC. Therefore, it can be inferred that IMEI and UserID are more commonly utilized to identify the users of the app compared with other device identifiers. Besides, the leakage of phone number shows a little difference. It can be seen that 9 $<app, \ Key>$ pairs transmit phone number, while the leakage times of phone number reaches 49453. After analyzing these 9 samples, we found that 44561 privacy leaks occurred in the same app. What's more, the app is the official application of a telecom service provider in China. Lastly, only two $<app, \ Key>$ pairs are found in our dataset to transmit email addresses, hence it seems that the email address has been well protected during its collection.

\subsection{Comparison of machine learning classifiers}
\begin{figure}[h]
    \scriptsize
    \centering
    \includegraphics[width=0.45\textwidth]{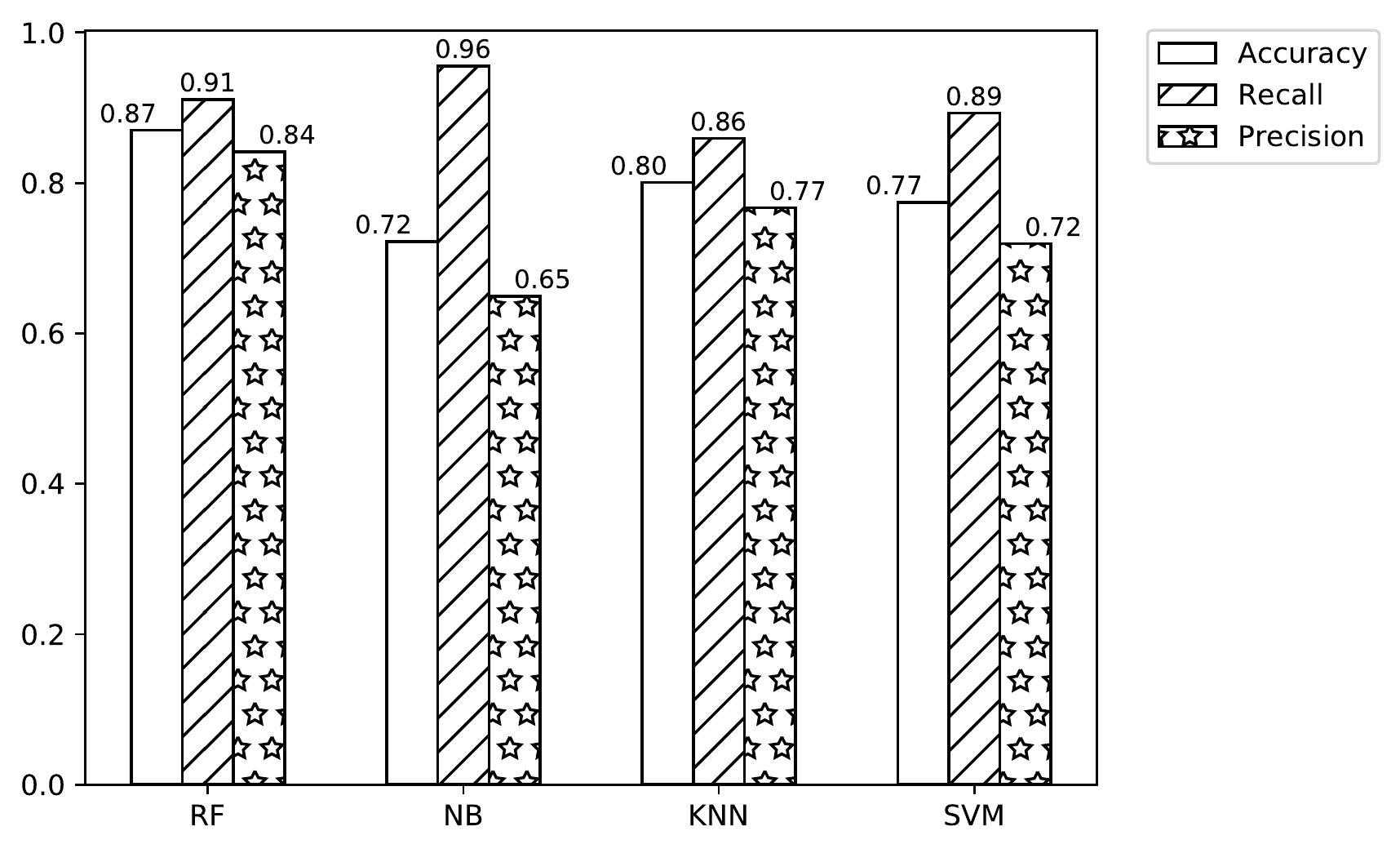}
    \caption{The performance comparison of different detectors.}
    \label{fig.5}
\end{figure}

To verify the impact of machine learning algorithms on the performance of the detector, four machine learning algorithms are applied to train the detector, including Random Forest (20 trees), SVM with Gaussian kernel, KNN, and Gaussian Naive Bayes. The labeled dataset is randomly divided into a training set and a testing set with a split ratio of 8:2. Afterward, the detectors are trained with the training set, and their performance is evaluated on the testing set. Three evaluation metrics in machine learning are used to evaluate the performance of the detectors, including precision, recall, and accuracy. After repeating the above process 10 times, the average performance of four detectors is provided in Fig.\ref{fig.5}. 

As illustrated in Fig.\ref{fig.5}, the detector based on Random Forest achieves the highest accuracy and precision, which are 87\% and 84\% respectively. Besides, the recall of the Random Forest detector reaches 91\%. The performance of KNN is slightly lefts behind Random Forest. Although Naive Bayes realizes the highest recall, its precision and accuracy are far lower than that of the other three detectors. SVM presents a similar performance as Naive Bayes. In view of the above results, the Random Forest detector is employed in all subsequent experiments.   

\subsection{Effectiveness of the proposed features}
The effectiveness of the proposed local features and global features is further analyzed in this Section. The performance of detectors trained on different feature sets is shown in Table \ref{tab3}. 
\begin{table}[h]
\scriptsize
\renewcommand\arraystretch{1.3} 
\setlength\tabcolsep{9pt}
\centering
\caption{The performance of detectors trained on different feature sets.}
\label{tab3}
\begin{tabular}{cccc}
\hline
Features        & Precision & Recall & Accuracy \\ \hline
Local Features  & 0.6718    & 0.7179 & 0.6870   \\
Global Features & 0.8265    & 0.8817 & 0.8475   \\
All Features    & 0.8322    & 0.9177 & 0.8703   \\ \hline
\end{tabular}
\end{table}

It can be seen from Table \ref{tab3} that the performance of the detector is the worst when only local features are used. A possible reason is that more than half of the samples involve one user only, thus the local features cannot distinguish PI from non-PI effectively. For instance, the keys that transmit IMEI and the package name are likely to have similar local features when all the app traffic comes from one user. On the contrary, those two keys would have different L-VRD when there is multiple users' traffic. In this case, the detector based on local features may realize better performance. Meanwhile, the detector based on global features shows a slightly lower performance than the detector based on whole features. Therefore, it can be inferred that the prediction of the $<app, \  Key>$ pair would be more reliable when sufficient global features can be extracted.

\subsection{Analysis on False Prediction}
As shown in the above experiments, the proposed detector has a false prediction rate of about 13\%. As stated earlier, the number of users may be a key factor affecting the prediction performance. In this section, we further analyze how this factor affects the performance of the detector. In the meantime, another factor, i.e. the source of false predictions, is investigated in detail. Finally, the prediction confidence is utilized to improve the prediction accuracy.

 (1) The number of users

\begin{figure}[h]
  \scriptsize
  \centering
  \includegraphics[width=0.4\textwidth]{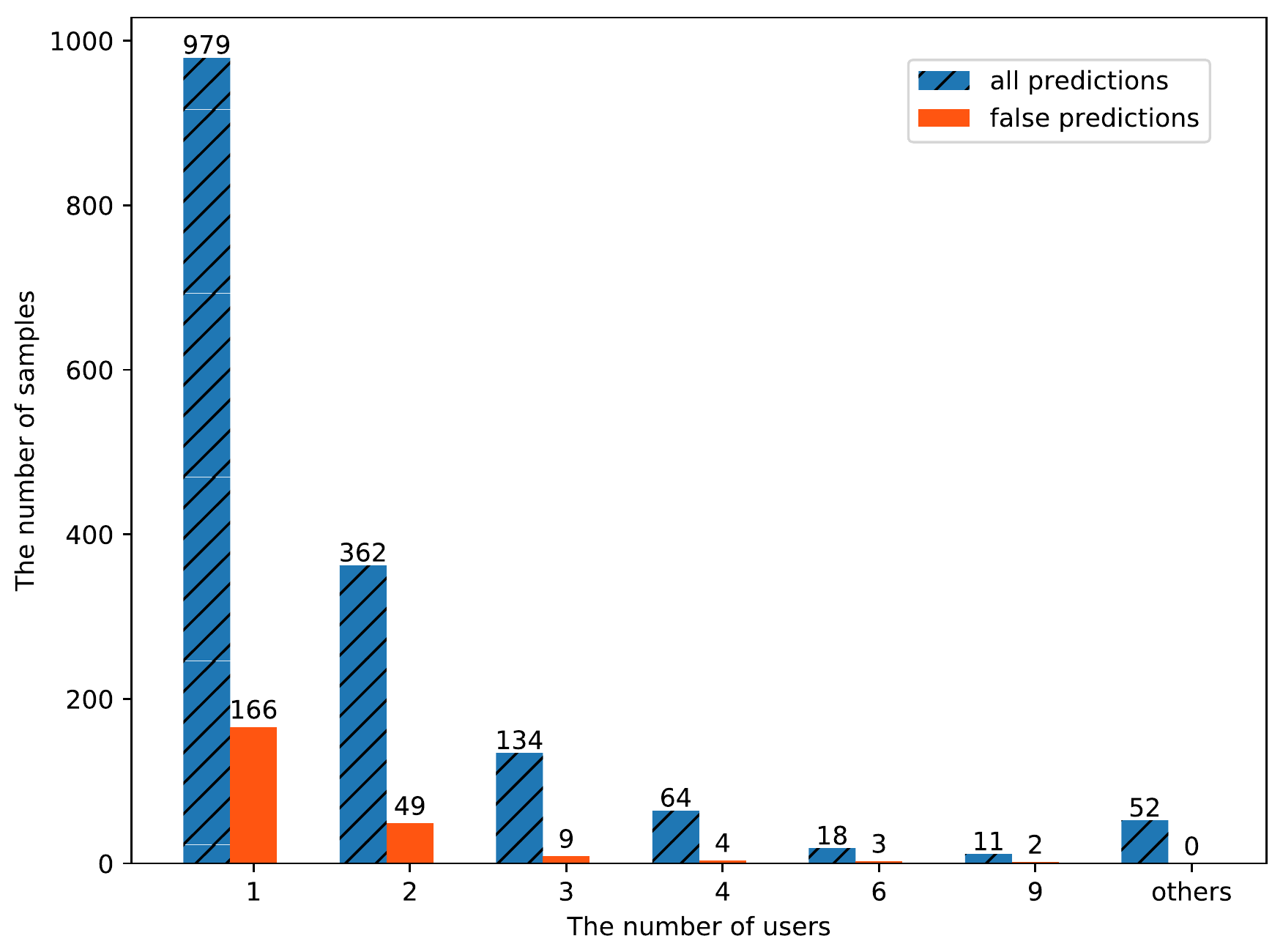}
  \caption{The distribution of the number of users for the false predictions.}
  \label{fig.6}
\end{figure}

To illustrate how the number of users affects the performance of the detector, the distribution of the number of users for the false predictions in 10 experiments is depicted in Fig.\ref{fig.6}. The category ``others" in Fig.\ref{fig.6} includes 11 types of number of users, i.e., 5, 7, 8, 10, 11, 13, 14, 17, 39, 41, 43.

As Fig.\ref{fig.6} shows, most false predictions result from the samples with a small number of users. When there are one or two users for a sample, the false prediction rate reaches 16.95\% and 13.53\%, respectively. With the increase in the number of users, the false predictions decrease significantly. Noted that there are no false predictions for the 52 samples with the category ``others". On the whole, only 18 false predictions are generated out of 279 predictions when the predicted samples have more than two users. In this case, the false prediction rate is reduced to 6.45\%. Therefore, the number of users of a sample is an important factor affecting prediction accuracy. More users exist for a sample, more reliable its prediction result is.

(2) Source of false predictions

\begin{figure}[h]
  \scriptsize
  \centering
  \includegraphics[width=0.4\textwidth]{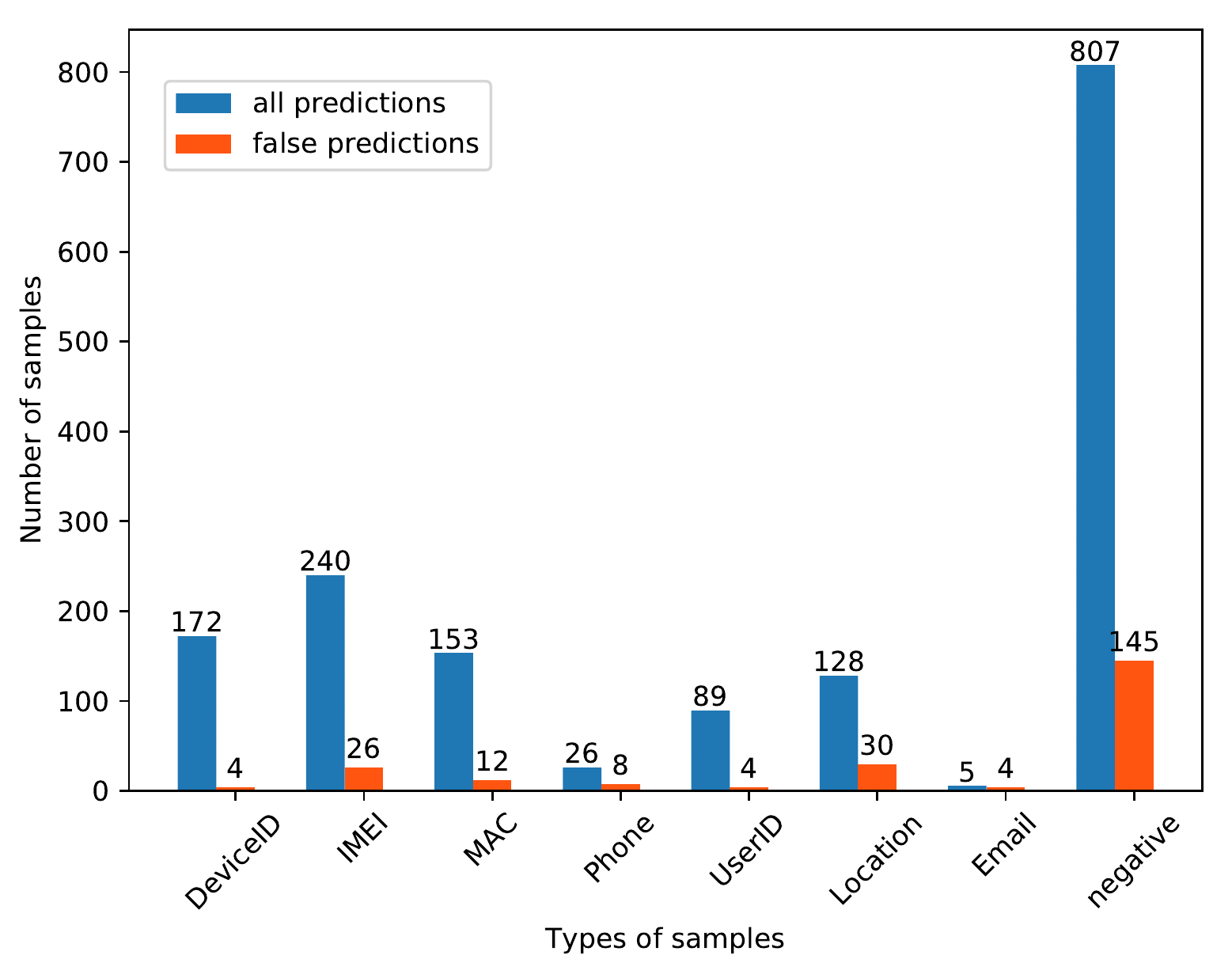}
  \caption{The type distribution of false predictions.}
  \label{fig.7}
\end{figure}

In 10 experiments, the true labels of all predicted samples and false predicted samples are collected. The type distribution of false predictions is shown in Fig.\ref{fig.7}.

It can be seen from Fig.\ref{fig.7} that the false positive is the main source of the false prediction. In the prediction of 807 negative samples, 145 samples are classified as positive incorrectly. In contrast, 88 false negatives are generated in the prediction of 808 positive samples. Moreover, the detector is expected to generate as few false positives as possible in many application scenarios. Therefore, a measure is taken in the following Section to improve the performance of the detector.

Additionally, more observations can be obtained from Fig.\ref{fig.7}. Firstly, our detector cannot identify the ``Email" samples effectively. One possible reason is that there are only two email samples in our dataset. Even worse, one sample involves one user and another sample only has four valid values. Another observation is that ``location" samples are more prone to be falsely predicted than the samples with other PI types. Compared with the other six kinds of PI, location has a different trait. That is, its value could change frequently when the user moves. This trait may cause it to be difficult to distinguish it from other app information. Therefore, more features can be added to identify location information separately in future work, such as the data format. 

(3) Prediction confidence

\begin{figure}[h]
  \scriptsize
  \centering
  \includegraphics[width=0.4\textwidth]{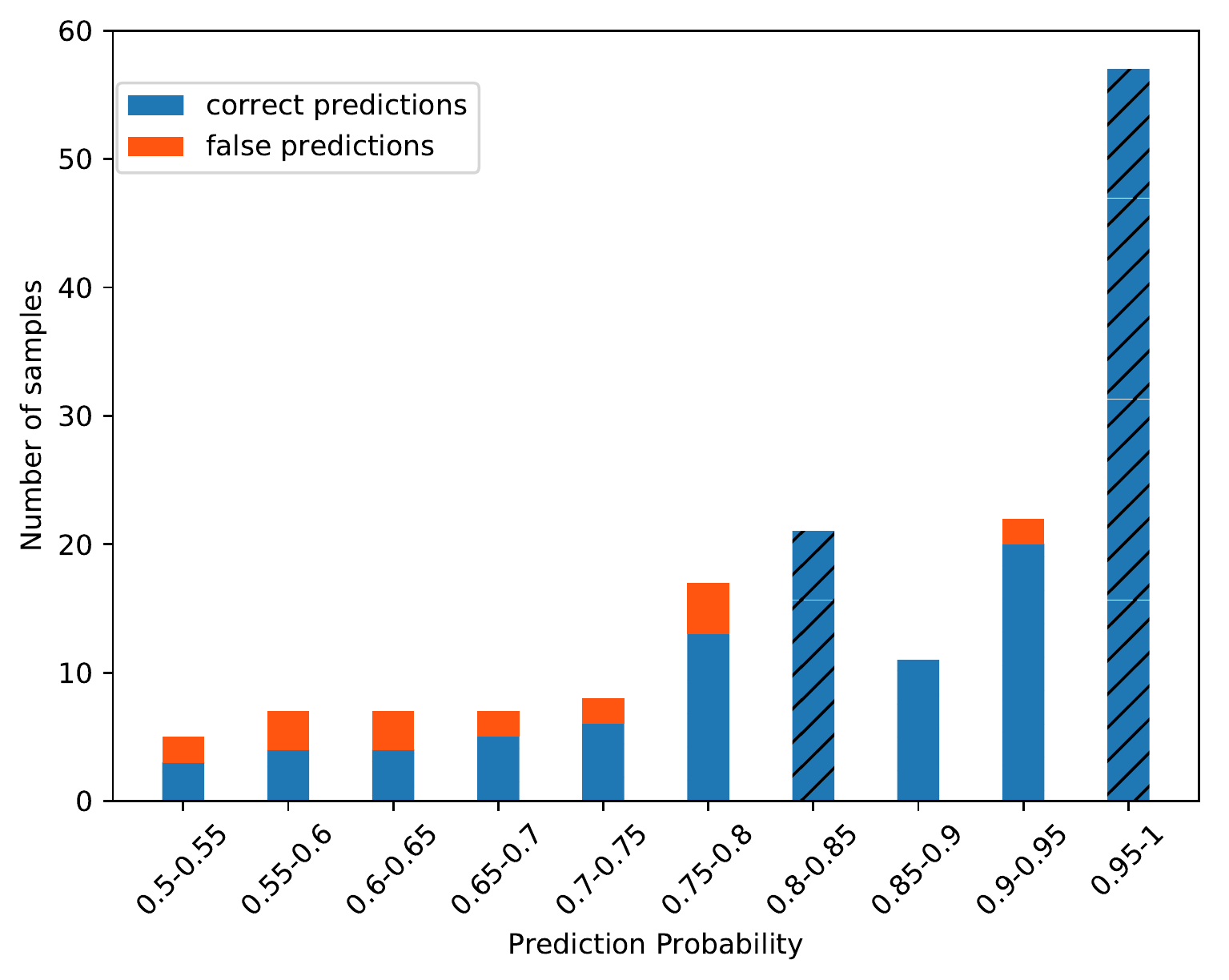}
  \caption{The prediction probability distribution.}
  \label{fig.8}
\end{figure}

During the prediction, the Random Forest detector can not only give the label but also provide the probability of this prediction. Fig.\ref{fig.8} illustrates the prediction probability of one random testing. 
  
It can be observed from Fig.\ref{fig.8} that the correct prediction usually has a high prediction probability. By contrast, the probability of a false prediction is likely to be below 0.75. Therefore, prediction confidence can be utilized to reduce false predictions. When the prediction probability is lower than the confidence, the detector would reject to give a prediction. Although this measure would decrease the recall of the detector, the precision can be improved significantly in practice. For example, when the confidence is set as 0.75, 7 of the original 10 false positives are rejected, and the precision of positive samples would be improved to 95.58\%. In the meantime, the recall of positive samples is 74.71\%. In total, 79.01\% of samples can be predicted.

\subsection{Prediction on unlabeled samples}
Based on the Random Forest detector given in Section 5.2, and the confidence is set as 0.75, the remaining 16160 unlabeled samples in our dataset are classified. Finally, a total of 909 samples are identified as PI-related $<app, Key>$ pairs. We manually check those 909 samples to validate their predictions. 

Since the value of PI is unknown and even has been obfuscated, the predictions cannot be verified objectively. Under this circumstance, conservative evaluation strategies are applied. The strategies are set as follows: (1) When the information transmitted by the $<app, \ Key>$ pair is evidently not PI, such as URLs or domain, then a false positive is generated. Contrarily, the prediction is correct for the $<app, \ Key>$ pair with PI-related values or its key has definite PI-related semantics. (2) If semantic information cannot be inferred from the $<app,\ Key>$ pair, the prediction is regarded as a false positive when the key carries different values for one user. Otherwise, the prediction of this $<app, \ Key>$ pair is uncertain. Strategy (2) is conservative since the PI that is similar to the location would be misjudged. Another example is that the case where a mobile device has two identifiers, such as two IMEIs. In this case, strategy (2) would also misjudge the prediction. Therefore, the following analysis would give a lower bound of the performance of our detector. According to the above strategies, the predictions of the 909 samples are listed in Table \ref{tab4}.
\begin{table}[h]
\scriptsize
\renewcommand\arraystretch{1.3} 
\setlength\tabcolsep{9pt}
\centering
\caption{The details of the 909 samples with positive prediction.}
\label{tab4}
\begin{tabular}{ccc}
\hline
Type                          & \# of \textless{}app, Key\textgreater{} & \# of apps \\ \hline
AndroidID                        & 30                                      & 27         \\
Device Information                     & 104                                     & 70         \\
IMEI                             & 13                                      & 13         \\
IP                               & 16                                      & 16         \\
Username                         & 1                                       & 1          \\
app Install Time                 & 1                                       & 1          \\
Location                         & 3                                       & 3          \\
ISP                              & 2                                       & 2          \\
Screen Size                      & 22                                      & 20         \\
Serial Number                    & 2                                       & 2          \\
app Download Market                  & 2                                       & 2          \\
Uncertain                        & 443                                     & /          \\ \hline
\multirow{2}{*}{Potential False Positives} & 142                                     & /          \\ \cline{2-3} 
                                 & 128                                     & /          \\ \hline
\end{tabular}
\end{table}

\begin{table*}[bp]
\scriptsize
\renewcommand\arraystretch{1.3} 
\setlength\tabcolsep{7pt}
\centering
\caption{The difference between the existing methods and ours}
\label{tab5}
\begin{threeparttable}
\begin{tabular}{ccccccc}
\hline
Work            & Features              & Model            & SSL & Obfuscation & Unknown types of PI&\# of Classifiers \\ \hline
Seeded approach\cite{Liu2015} & String Pattern       & String Matching  & N   & N      & N    & /                 \\
Recon\cite{Recon2016}              & Bag-of-Words         & Machine-Learning & Y   & N    & N       & $10^3$               \\
Antshield\cite{Antshield2018}        & Bag-of-Words         & Machine-Learning & Y   & N    & N       & $10^2$               \\
Our Method      & Statistical Features & Machine-Learning & N\tnote{*}   & Y     & Y      & 1                 \\ \hline
\end{tabular}
\begin{tablenotes}
  \item[*] As Recon and Antshield did, our method could integrate MITM to deal with encrypted traffic in future work, .
\end{tablenotes}
\end{threeparttable}
\end{table*}

As shown in Table \ref{tab4}, various PI has been identified, which is far beyond the six types of PII in the labeled dataset. It proves that our detector is capable of discovering privacy leaks of unknown types. Among these privacy leaks, device information and AndroidID are the two most frequently leaked information. The device information includes the phone model, the phone brand, and CPU info, etc. In addition to these common PI, it can be seen that some user behavior-related PI is also collected by apps, such as the app install time and the market that downloads the app. Note that those collections could happen without user consent and awareness. The proposed detector in this paper could assist in detecting such privacy leaks.

Besides, other interesting points are further found during our manually checking process. Firstly, for several $<app, \ Key>$ pairs, their keys contain the string ``imei\_md5", and their values are unique to each user. Therefore, it can be inferred reasonably that they are used to transmit the obfuscated IMEI. As for the 443 uncertain samples, all of their values appear to be obfuscated strings. Besides, most $<app, \ Key>$ pairs have the string ``**id" in their keys. Hence we speculate that these keys are likely to transmit certain user-related identifiers. 

Lastly, 270 samples are judged as false positives. Among those samples, 142 samples have a unique value for each user. In the meantime, 134 samples out of the 142 samples have only one user. We believe some of those false positives can be eliminated when more users' traffic is added. The rest 128 false positives are judged by the strategy (2). As aforementioned, those judgments are conservative. Moreover, we do find that some $<app, \ Key>$ pairs have at most two values for each user.

All in all, the above analysis demonstrates that the proposed detector can automatically identify privacy leaks in the traffic, including the unknown types of privacy leaks and the privacy leaks with obfuscation.

\subsection{Comparison with other works}

In this Section, the proposed detector is compared with three existing traffic monitoring-based detection methods based on our labeled dataset. The difference between those three works and ours is summarized in Table \ref{tab5}. Compared with the existing three methods, the proposed detector can detect the PI with obfuscation and unknown types. In terms of the number of classifiers, our method builds one general classifier, while Recon and Antshield train a classifier for each domain and app, respectively. Besides, with one prediction, our method can identify the privacy leaks in the traffic through simple string matching. However, Recon and Antshield have to classify each traffic flow by the classifiers. Therefore, less overhead would be introduced by our method than Recon and Antshield. 

Table \ref{tab6} gives the performance comparison of those methods.  It can be seen from Table \ref{tab6} that the simplest method, i.e. Seeded approach, has the worst performance. The Seeded approach only identifies 3 positive samples correctly. Meanwhile, it generates 26 false positives. Recon achieves the best performance among the existing three methods. Compared with Recon, our detector has the same recall with it. However, our detector achieves higher precision than Recon. In another way, the false positives generated by our detector is 20 less than that generated by Recon. In conclusion, the performance of the proposed detector in this paper is better than the existing methods. 

\begin{table}[h]
\scriptsize
\renewcommand\arraystretch{1.3} 
\setlength\tabcolsep{2pt}
\centering
\caption{The performance comparison with other works.}
\label{tab6}
\begin{tabular}{cccccc}
\hline
Work                                                       & False Positive & False Negative & Precision & Recall  & F1-measure \\ \hline
Recon\cite{Recon2016}                                                      & 28 & 12 & 71.13\%   & 85.18\% & 77.52\%    \\
Antshield\cite{Antshield2018}                                                  & 30 & 19 & 67.39\%   & 76.54\% & 71.67\%    \\
\begin{tabular}[c]{@{}c@{}}Seeded \\ approach\cite{Liu2015}\end{tabular} & 26 & 78 & /         & /       & /          \\
Our Method                                                 & 8  & 12 & 89.61\%   & 85.18\% & 87.34\%    \\ \hline
\end{tabular}
\end{table}

\section{Conclusion}
To discover privacy leaks in mobile traffic, a novel detection method based on traffic monitoring is presented in this paper. The proposed method utilizes statistical features to capture the occurrence patterns of personal information in the traffic. Based on machine learning, the detector could identify more potential personal information in the traffic. Compared with the existing methods, the proposed method can identify unknown types of personal information. Besides, the proposed method is resistant to obfuscation technology. Finally, the experimental results show that the proposed method could achieve better performance than the existing methods.

\section*{Acknowledgements}
This work was supported by the National Key Research and Development Program of China (2018YFB0204301) and The Science and Technology Innovation Program of Hunan Province(2020RC2047).


\begin{thebibliography}{99}

%% \bibitem[Author(year)]{label}{}
\bibitem{finonline} Number of Mobile app Downloads in 2021/2022: Statistics, Current Trends, and Predictions. https://financesonline.com/number-of-mobile-app-downloads/. 
\bibitem{chadza2017} Chadza. T, Aparicio-Navarro. F, Kyriakopoulos. K, Chambers. J. A Look into the Information Your Smartphone Leaks[C]// International Symposium on Networks, Computers and Communications, pp.1-6, 2017. doi:10.1109/isncc.2017.8072022. 
\bibitem{He2019} Yongzhong He, Xiaojuan zhao, Chao Wang. Privacy Mining of Large-scale Mobile Usage Data[C]// IEEE International Conference on Power, Intelligent Computing and Systems, pp. 81-86, 2019. doi:10.1109/ICPICS47731.2019.8942559.
\bibitem{facebook} Facebook data privacy scandal: A cheat sheet. https://www.techrepublic.com/article/facebook-data-privacy-scandal-a-cheat-sheet/.
\bibitem{Rela} Rela, a Chinese Lesbian Dating app, Exposed 5 Million User Profiles. https://techcrunch.com/2019/03/27/rela-data-exposed/.
\bibitem{Ren2018} Jingjing Ren, Martina Lindorfer, Daniel J. Dubois, et al. Bug Fixes, Improvements,... and Privacy Leaks, A Longitudinal Study of PII Leaks Across Android app Version[C]// Network and Distributed System Security Symposium, 2018. doi:10.14722/ndss.2018.23159.
\bibitem{Egele2011} Manuel Egele, Christopher Kruegel, Engin Kirda, et al. PiOS: Detecting Privacy Leaks in iOS applications[C]// Proceedings of the Network and Distributed System Security Symposium, 2011. 
\bibitem{AndroidLeaks} Clint Gibler, Jonathan Crussell, Jeremy Erickson, et al. AndroidLeaks: Automatically Detecting Potential Privacy Leaks in Android applications on A Large Scale[C]// International Conference on Trust and Trustworthy Computing, pp.291-307, 2012. doi:10.1007/978-3-642-30921-2\_17.
\bibitem{MobileappScrutinator} J. p. Achara, V. Roca, C. Castelluccia, et al. MobileappScrutinator: A Simple Yet Efficient Dynamic Analysis approach for Detecting Privacy Leaks across Mobile OSs[J]. Arxiv, abs/1605.08357, 2016.
\bibitem{AGRIGENTO} A. Continella, Y. Fratantonio, M Lindorfer, et al. Obfuscation-Resilient Privacy Leak Detection for Mobile apps Through Differential Analysis[C]// Proceedings of the Network and Distributed System Security Symposium, 2017. doi:10.14722/ndss.2017.23465.
\bibitem{Antmonitor2015} Anastasia Shuba, Anh Le, Minas Gjoka, et al. AntMonitor: Network Traffic Monitoring and Real-Time Prevention of Privacy Leaks in Mobile Devices[C]// Proceedings of the 2015 Workshop on Wireless of the Students, by the Students and for the Students, pp.25-27, 2015. doi:10.1145/2801694.2801707.
\bibitem{Recon2016} Jingjing Ren, Ashwin Rao, Martina Lindorfer, et al. ReCon: Revealing and Controlling Privacy Leaks in Mobile Network Traffic[C]// Proceedings of the 14th Annual International Conference on Mobile Systems, applications, and Services, pp.361-374, 2016. doi:10.1145/2906388.2906392.
\bibitem{Antshield2018} Anastasia Shuba, Evita Bakopoulou, Mehrabadi Asgari, et al. AntShield: On-Device Detection of Personal Information Exposure[J]. Arxiv, abs/1803.01261, 2018.
\bibitem{zhang2021} Tengfei Zhang, Shunzheng Yu. Research Prospects of User Information Detection from Encrypted Traffic of Mobile Devices[J]. Journal on Communications, 42(2), pp.154-167, 2021.
\bibitem{GB2020} The Standardization Administration of China. Information Security Technology - Personal Information Security Specification. The Chinese National Standard GB/T 35273-2020, 2020.
\bibitem{LeakDoctor} Xiaolei Wang, Andrea Continella, Yuexiang Yang, et al. LeakDoctor: Toward Automatically Diagnosing Privacy Leaks in Mobile applications[J]// Proceedings of the ACM on Interactive, Mobile, Wearable and Ubiquitous Technologies, 3(1), pp.1-25, 2019. doi:10.1145/3314415. 
\bibitem{BATLeaks} Yuefei Zhao, Longtao He, Zhoujun Li, et al. Large-scale Detection of Privacy Leaks for BAT Browsers Extensions in China[C]// International Symposium on Theoretical Aspects of Software Engineering, pp.57-64, 2019. 
\bibitem{FlowDroid2014} Steven Arzt, Siegfried Rasthofer, Christian Fritz, et al. FlowDroid: Precise Context, Flow, Field, Object-sensitive and Lifecycle-aware Taint Analysis for Android apps[C]// Proceedings of the 35th ACM SIGPLAN Conference on Programming Language Design and Implementation, pp.259-269, 2014. doi:10.1145/2594291.2594299.
\bibitem{Nattanon2020} Nattanon Wongwiwatchai, Phannawhat Pongkham, Kunwadee Sripanidkulchai, et al. Comprehensive Detection of Vulnerable Personal Information Leaks in Android applications[C]// IEEE INFOCOM - IEEE Conference on Computer Communications Workshops, 2020. doi:10.1109/INFOCOMWKSHPS50562.2020.9163043.
\bibitem{ClueFinder} Yuhong Nan, Zhemin Yang, Xiaofeng Wang, et al. Finding Clues for Your Secrets: Semantics-Driven, Learning-Based Privacy Discovery in Mobile apps[C]// Proceedings of the 24th Network and Distributed System Security Symposium, 2018. doi:10.14722/ndss.2018.23099.
\bibitem{TaintDroid2014} William Enck, Peter Gilbert, Byung-Gon Chun, et al. TaintDroid: An Information Flow Tracking System for Real-Time Privacy Monitoring on Smartphones[J]. Communications of the ACM, 57(3), pp.99-106, 2014. doi:10.1145/2494522. 
\bibitem{privacycapsules}Raul Herbster, Scott DellaTorre, Peter Druschel, et al. Privacy Capsules: Preventing Information Leaks by Mobile apps[C]// Proceedings of the 14th Annual International Conference on Mobile Systems, and Services, pp.399-411, 2016. doi:10.1145/2906388.2906409. 
\bibitem{iABC} A. J. Bhatt, Chetna Gupta, Sangeeta Mittal, et al. iABC: Towards a Hybrid Framework for Analyzing and Classifying Behaviour of iOS applications Using Static and Dynamic Analysis[J]. Journal of Information Security and applications, 41, pp.144-158, 2018. 
\bibitem{He2018} Yongzhong He, Binghui Hu, Zhen Han. Dynamic Privacy Leakage Analysis of Android Third-party Libraries[C]// International Conference on Data Intelligence and Security, pp.275-280, 2018. doi:10.1109/ICDIS.2018.00051. 
\bibitem{PrivacyGuard} Yihang Song, Urs Hengartner. PrivacyGuard: A VPN-based Platform to Detect Information Leakage on Android Devices[C]// Proceedings of the Annual ACM CCS Workshop on Security and Privacy in Smartphones and Mobile Devices, pp.15-26, 2015. doi:10.1145/2808117.2808120.
\bibitem{Haystack} Abbas Razaghpanah, Narseo Vallina-Rodriguez, Srikanth Sundaresan, et al. Haystack: A Multi-Purpose Mobile Vantage Point in User Space[J]. arxiv, abs/1510.01419v3, 2015. 
\bibitem{Liu2015} Yabing Liu, Han Hee Song, Ignacio Bermudez, et al. Identifying Personal Information in Internet Traffic[C]// Proceedings of the ACM on Conference on Online Social Networks, pp.59-70, 2015. doi:10.1145/2817946.2817947.
\bibitem{Shen2019} Liang Shen, Xin Wang, Shuhui Chen. Structural Signature Extraction Method for Mobile application Recognition[J]. Journal of Computer applications, 2020, 40(4):1109-1114.  
\bibitem{ours2019} Shuhui Chen, Shuang Zhao, Biao Han, et al. Investigating and Revealing Privacy Leaks in Mobile application Traffic[C]// 2019 Wireless Days, pp.1-4, 2019. doi:10.1109/WD.2019.8734246. 
\bibitem{Netlog} Xin Wang, Shuhui Chen, Jinshu Su. Automatic Mobile app Identification From Encrypted Traffic With Hybrid Neural Networks[J]. IEEE Access, 8, pp.182065-182077, 2020. doi:10.1109/ACCESS.2020.3029190. 



\end{thebibliography}
\end{document}